%% file: ms.tex
\documentclass[conference]{IEEEtran}

\usepackage{array, calc, color, multicol}
\usepackage{graphics, epstopdf, graphicx, epsfig, epsf, subfigure}
\usepackage{amssymb, amsfonts, amsmath, times}
\usepackage{multicol,balance, verbatim}
\usepackage{textcomp, mathrsfs, stmaryrd, setspace}
\usepackage{amssymb}
\usepackage{algorithm}
\usepackage[noend]{algpseudocode}
\usepackage{url}

\newcommand{\ie}{{\em i.e., }}

\newcommand{\eg}{{\em e.g., }}

\newtheorem{example}{Example}

\DeclareGraphicsExtensions{.eps, .pdf, .png, .jpg}   

\newcommand{\Aset}{\mathcal{A}}

\newcommand{\Nset}{\mathcal{N}}

\IEEEoverridecommandlockouts

\makeatletter
\newcommand{\oset}[2]{%
{\mathop{#2}\limits^{\vbox to -.5\ex@{\kern-\tw@\ex@
\hbox{\scriptsize #1}\vss}}}}
\makeatother

\begin{document}

\title{AACT: Application-Aware Cooperative Time Allocation for Internet of Things}

\author{Raheleh Khodabakhsh and Hulya Seferoglu\\
{ ECE Department, University of Illinois at Chicago}\\
{ \tt rkhoda2@uic.edu, hulya@uic.edu}
}

\maketitle

\input{abstract}
\input{introduction}

\input{system}

\input{problem_and_rhc}

\input{simulation}

\input{related}
\input{conclusion}
\input{refs}

\end{document}

%% file: abstract.tex
\begin{abstract}
As the number of Internet of Things (IoT) devices keeps increasing, data is required to be communicated and processed by these devices at unprecedented rates. Cooperation among wireless devices by exploiting Device-to-Device (D2D) connections is promising, where aggregated resources in a cooperative setup can be utilized by all devices, which would increase the total utility of the setup. In this paper, we focus on the resource allocation problem for cooperating IoT devices with multiple heterogeneous applications. In particular, we develop Application-Aware Cooperative Time allocation (AACT) framework, which optimizes the time that each application utilizes the aggregated system resources by taking into account heterogeneous device constraints and application requirements. AACT is grounded on the concept of Rolling Horizon Control (RHC) where decisions are made by iteratively solving a convex optimization problem over a moving control window of estimated system parameters. The simulation results demonstrate significant performance gains. 
\end{abstract}

%% file: introduction.tex
\section{Introduction}

\begin{figure}[!t]
\vspace{-5pt}
\centering
\includegraphics[height=5cm]{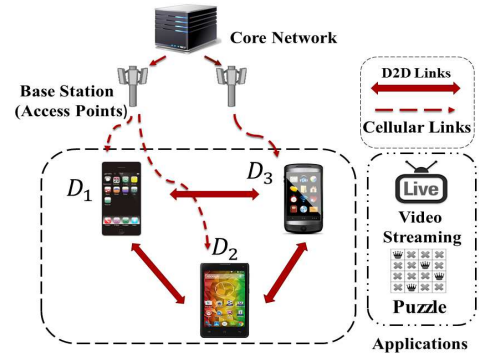}
\vspace{-5pt}
\caption{Example scenario of cooperating devices. Devices (i) have  cellular connections, and (ii) D2D connections, which make direct communication among devices possible without going through an auxiliary device such as a base station. The cooperating devices have two applications of interest; video streaming, and solving an N-queen puzzle problem.
}
\vspace{-15pt}
\label{fig:example}
\end{figure}

As the number of Internet of Things (IoT) devices including smartphones, wireless sensors, smart meters, and health monitoring devices keep increasing \cite{cisco_15_20}, data is required to be communicated and processed by these devices at unprecedented rates. Device-to-device (D2D) networking is a major enabling technology for future networks, allowing IoT and wireless devices to communicate directly to each other without going through an auxiliary device, Fig.~\ref{fig:example}.

D2D connections with very high rates (up to 250 Mbps using Wi-Fi Direct \cite{wifi_alliance})
 make cooperation among IoT devices promising to effectively utilize scarce IoT resources such as computing power, energy, and cellular bandwidth. In this context, our goal is to develop an application-aware cooperation framework in D2D networks.

Cooperation among wireless devices by exploiting D2D connections is getting increasing interest, where aggregated resources in a cooperative setup can be utilized by all devices, which would increase the total utility of the cooperative setup. For example, cellular links in Fig.~\ref{fig:example} cooperatively used in \cite{keller2012microcast} to improve the throughput of each device. Similarly, such a cooperative system is used for cellular offloading \cite{andreev2014cellular,andreev2015analyzing}, cooperative computation \cite{singh2016energy,cao2016share}, and energy efficiency \cite{ramadan2008implementation,shen2016device,singh2016energy}. However, existing work focuses on one specific application usually, \eg video streaming \cite{keller2012microcast,gong2016design,duong2015energy} or computation \cite{singh2016energy,cao2016share}. On the other hand, IoT and mobile devices typically support multiple applications simultaneously with different resource requirements. The key question in this context is how to design a resource allocation mechanisms for cooperating IoT devices using D2D links by taking into account application specific requirements. The next example illustrates our point.

\begin{example} \label{ex:intro}
Let us consider an example topology in Fig.~\ref{fig:example}, where three devices ($D_1$, $D_2$, $D_3$) are connected to each other via D2D links. Each device is also connected to the Internet via cellular links, where the downlink bandwidths are $1Mbps$, $2Mbps$, and $2Mbps$ for $D_1$, $D_2$, and $D_3$, respectively. We assume that $D_1$ is Nexus 6P smartphone with $4x1.55GHz$ \cite{Google6p}, while $D_2$ and $D_3$ are Nexus 5 smartphone with $2.3GHz$ computation power \cite{Google5}, the cellular and D2D links operate simultaneously (as they can use different parts of the spectrum), and D2D links are high; $100Mbps$.

Assume that there are two applications: (i) video streaming; intensive in terms of cellular downlink rate, and (ii) computation application; intensive in terms of computation power (\eg solving N-queen puzzle, face detection, etc.). $D_1$ is interested in video streaming, while $D_2$ and $D_3$ are interested in the computation application.

In this setup, if $D_2$ and $D_3$ help $D_1$ to stream the video, the streaming rate of $D_1$ increases to $5Mbps$ from $1Mbps$, which is significant. However, helping $D_1$ may delay the processing of the computation application at $D_2$ and $D_3$, which is not ideal for $D_2$ and $D_3$. Thus, a resource allocation mechanism should be developed so that cellular bandwidth and computation resources of all devices can be efficiently utilized by video streaming and computation applications by taking into account their application specific requirements.\footnote{Note that a naive solution would be isolating applications so that $D_1$ streams video only, while $D_2$ and $D_3$ process the computation application. However, such an isolation is suboptimal as $D_1$ is stronger in terms of computation, while $D_2$ and $D_3$ are stronger in terms of cellular bandwidth. Thus, their resources are complementary and we can get the optimal result if these devices cooperate and help each other.}
\hfill $\Box$
\end{example}

The primary challenge in a cooperative system with multiple applications is to define a unifying performance metric. For example, if we only have a video streaming application in Example~\ref{ex:intro}, the performance metric could be download rate of the video, and we can develop a cooperative resource allocation mechanism to maximize the download rate. Similarly, the performance metric of computation application in Example~\ref{ex:intro} could be computation completion time, which can be minimized. The crucial question when there exists multiple applications with different expectations from the system (\eg download rate or computation completion time) is to define a unifying performance metric. In this paper, we define this metric as time, and develop a resource allocation framework that optimizes the time that each application utilizes the cooperative system resources by taking into account application requirements.

In this paper, we develop Application-Aware Cooperative Time allocation (AACT) for IoT devices. We first model the time allocation problem as a utility optimization problem, which takes into account application requirements, device resources, and resource consumption over time. For example, battery consumption over time in each device is modeled as a linear function of other resource consumptions such as computing power or cellular bandwidth. We solve our network utility maximization problem using Receding Horizon Control (RHC), which gives our AACT algorithm as a solution. AACT makes time allocation decisions by taking into account heterogeneous application requirements and device constraints, and by iteratively solving a convex optimization problem over a moving control window of estimated system parameters. Finally, we develop a distributed version of AACT, which is AACT-distributed. The following are key contributions of this work:


\begin{itemize}
\item We consider the setup of cooperative IoT devices with multiple applications running simultaneously. In this setup, our goal is to develop a resource allocation mechanism by taking into account the requirements of diverse applications, and constraints of heterogeneous IoT devices. The crucial question when there exists multiple applications with different expectations from the system is to define a unifying performance metric. In this paper, we define this metric as time.

\item We develop a resource allocation framework that optimizes the time that each application utilizes the cooperative system resources. We model the time allocation problem as a utility optimization problem, which takes into account application requirements, device resources, and resource consumption over time. For example, battery consumption over time in each device is modeled as a linear function of other resource consumptions such as computing power and cellular bandwidth.

\item We solve our network utility maximization problem using Receding Horizon Control (RHC), which gives our AACT algorithm as a solution. AACT makes time allocation decisions by taking into account heterogeneous application requirements and device constraints in a time horizon. We also develop a distributed version of AACT, which is AACT-distributed.

\item We evaluate our proposed AACT and AACT-distributed algorithms for heterogeneous resources including cellular bandwidth and computing power, varying number of cooperating devices, and different applications. The simulation results shows AACT and AACT-distributed significantly improves completion time of applications and energy consumption of devices as compared to baselines.
\end{itemize}

The structure of the rest of this paper is as follows. Section \ref{sec:model} presents system model. Section \ref{sec:form} presents our Application-Aware Cooperative Time allocation (AACT) framework for IoT. Section \ref{sec:sims} evaluates the performance of our framework. Section \ref{sec:related} presents related work. Section \ref{sec:conclusion} concludes the paper.

%% file: system.tex
\section{\label{sec:model} System Model} 

We consider a setup with $N$ cooperative IoT devices, where $\Nset$ is the set of devices in our system with $N=|\Nset|$. These devices are withing close proximity of each other, so they are in the same transmission range and connected to each other (and form a clique topology) via D2D links (such as Wi-Fi Direct or Bluetooth). These devices are connected to the Internet via their cellular links.\footnote{We note that extension of our work to Wi-Fi connected devices to access the Internet is straightforward, so we omit this part in this paper for brevity.} We assume that cellular and D2D links use the different parts of the spectrum, so they can operate simultaneously. The example topology for $N=3$ devices is shown in Fig.~\ref{fig:example}. 

We consider a slotted system, where $t \in \mathcal{T}$ indicates the beginning of a time slot $t$, and $\mathcal{T}$ is the set of time slots beginning from $0$ to $T$ (noting that $T = |\mathcal{T}|$), which captures the time frame of interest in the system.

At slot $t$, cellular bandwidth of device $i$ is $\bar{w}_{i}^{t}$, which varies over time. We do not assume any channel model distribution in our problem formulation as our algorithm adapts itself to the observed state of channel conditions. In particular, we represent the amount of cellular downlink usage by device $i$ as $w_{i}^{t}$, which should be less than the observed cellular bandwidth; \ie 
\begin{align}\label{eq:cell_downlink_constraint}
w_{i}^{t} \leq \bar{w}_{i}^{t}
\end{align}
should be satisfied. 

On the other hand, we assume that D2D transmission rates are sufficiently high and not a bottleneck of the system model. This assumption is realistic as D2D transmission rates can be as high as $250Mbps$ in a close proximity   using Wi-Fi Direct \cite{wifi_alliance}. Furthermore, this assumption helps us better focus on scarce IoT resources such as cellular downlink rate, computation power, and energy. 

The computation power of device $i$ available at time $t$ is represented by $\bar{c}_{i}^{t}$, which varies over time, because device $i$ may process background (possibly operating system related) applications that we cannot control. Due to these uncontrolled applications, $\bar{c}_{i}^{t}$ varies over time. We assume that $c_{i}^{t}$ is the amount of computation power (in cycles/sec) that device $i$ utilizes, hence 
\begin{align}\label{eq:cell_cpu_constraint}
c_{i}^{t} \leq \bar{c}_{i}^{t}
\end{align}
should be satisfied. 

The amount of battery that device $i$ has at slot $t$ is represented by $b_i^t$, which takes a large value if there is a continuous power supply. On the other hand, $b_i^t$ decreases over time depending on how much device $i$ is utilized at slot $t$. In particular, if cellular links are used for downloading data for an application, the battery consumption will be related to $\gamma_{w}w_{i}^{t}$, where $w_{i}^{t}$ is the amount cellular bandwidth usage at time $t$ (as described earlier in this section), and $\gamma_{w}$ is the amount of battery consumption for every downloaded bits per second. Similarly, the amount of battery consumption due to processing an application is $\gamma_{c}c_{i}^{t}$, where $c_{i}^{t}$ is the amount of computation power and $\gamma_{c}$ is the amount of battery consumption per each cycle/sec. As a summary, the battery consumption from slot $t$ to $t+1$ varies according to:
\begin{align} \label{eq:battery_evol}
b_{i}^{t+1}= \alpha_i^t b_{i}^{t}-\gamma_{c}c_{i}^{t}-\gamma_{w}w_{i}^{t} \quad \forall i \in \mathcal{N},t\in \mathcal{T},
\end{align} where $\alpha_i^t$ captures the battery consumption due to background applications. 

We assume that $\Aset$ is the set of applications (tasks) in the system, where $A$ is the number of applications , \ie  $A = |\Aset|$. At least one device in the cooperative system is interested in an application. For example, devices $i$ and $k$ from $\Nset$ may be interested in application $j \in \Aset$. Although not all devices are interested in all applications, they will cooperate to handle all applications.  We note that the origin of each application file could be mobile devices themselves, base station, or the core network. Our framework is generic enough to capture all these scenarios. 

Each application $j\in \mathcal{A}$ has a certain resource requirements. For example, $a_{j}^{c}$ is the computing power requirement of application $j$, while $a_{j}^{w}$ is the cellular bandwidth requirement.

%% file: problem_and_rhc.tex
\section{\label{sec:form} AACT: Application-Aware Cooperative Time Allocation for Internet of Things}

In this section, we formulate and design our Application-Aware Cooperative Time allocation (AACT) framework for Internet of Things. Our first step is the Network Utility Maximization (NUM) formulation of the problem. Next, we solve our network utility maximization problem using Receding Horizon Control (RHC), which gives our AACT algorithm as a solution. Finally, we develop a distributed version of AACT, which is AACT-distributed.



\subsection{Network Utility Maximization}
The objective is to obtain an optimal policy which will efficiently distribute the tasks/applications across the mobile devices considering the heterogeneity of the devices and applications. As we mentioned earlier, the primary challenge in a cooperative system with multiple applications is to define a unifying performance metric. In this paper, we define this metric as time, and develop a resource allocation framework that optimizes the time that each application utilizes the cooperative system resources by taking into account application requirements. In particular, $\eta_{j}^{t}$  represents the percentage of the time spent on application $j$ across all mobile devices during time slot $t$. We associate $\eta_{j}^{t}$ to a utility function $U_{j}(.)$, where $U_{j}(\eta_j^t)$ is the utility of the system when application $j$ uses available resources $\eta_j^t$ percentage of time.\footnote{We assume that the utility function $U_{j}(.)$ is a continuously differentiable and concave function of $\eta_{j}^{t}$ for convergence and stability purposes.} Our utility maximization problem is expressed as; 
\begin{align} \label{eq:opt}
\max_{\boldsymbol {{\eta}}} & \sum_{j \in \mathcal{A}, t\in \mathcal{T}}U_{j}({\eta_{j}}^{t}) \nonumber \\
\mbox{s.t. } &   \sum_{j}\eta_{i,j}^{t}a_{j}^{w} \leq w_{i}^{t}, \quad \forall i \in \mathcal{N},t\in \mathcal{T} \nonumber \\ 
& \sum_{j}\eta_{i,j}^{t}a_{j}^{c} \leq c_{i}^{t}, \quad \forall i \in \mathcal{N},t\in \mathcal{T} \nonumber \\
& \eta_{j}^{t}\leq \sum_{i\in \mathcal{N}}\eta_{i,j}^{t}\quad \forall j \in \mathcal{A},t \in \mathcal{T} \nonumber \\
& \sum_{j}\eta_{i,j}^{t}\leq 1\quad \forall i \in \mathcal{N},t\in \mathcal{T} \nonumber \\
& (\ref{eq:cell_downlink_constraint}), (\ref{eq:cell_cpu_constraint}), (\ref{eq:battery_evol}),  
\end{align} where $\eta_{i,j}^{t}$ is the percentage of time device $i$ spends on application $j$ at slot $t$. The first constraint in (\ref{eq:opt}) states that the total amount of cellular bandwidth usage by all applications at device $i$ should not exceed the available cellular bandwidth at slot $t$. Similarly, the second constraint states that the total amount of computing power  usage by all applications at device $i$ should not exceed the available computing power. The third constraint relates $\eta_j^t$ and $\eta_{i,j}^{t}$. The fourth constraint states that the total percentage of time that at each device uses cannot exceed $1$. The final constraints of our utility maximization problem are the device constraints; (\ref{eq:cell_downlink_constraint}), (\ref{eq:cell_cpu_constraint}), (\ref{eq:battery_evol}). Next, we will present our approach for solving (\ref{eq:opt}). 

\subsection{ACCT via Receding Horizon Control (RHC)}

In this section, we develop our AACT algorithm using an online receding horizon control (RHC) approach (also known as model predictive control (MPC)) by iteratively solving the optimization problem in (\ref{eq:opt}). 

Let $\boldsymbol x^{t}$ be the set of all the control variables and battery level, \ie $\boldsymbol x^{t} = \{ \eta_{j}^{t}, c_{i}^{t}, w_{i}^{t}, \eta_{i,j}^{t}, b_{i}^{t} \}_{\forall i \in \Nset, j \in \Aset}$ at time slot $t$.  Moreover, let $\boldsymbol z^{t:t+\omega}$ denote the set of all estimated system parameters over a prediction window size of $\omega$ such as $\boldsymbol z^{t:t+\omega} = \{ \bar{c}_{i}^{t:t+\omega}, \bar{w}_{i}^{t:t+\omega} \}_{ \forall i \in \Nset}$.  

The decisions in AACT are made as follows. For $t=0$, we set $b_{i}^{t}=b_{i}^{0}$, $\forall i \in \Nset$. At each time slot $t$, we solve the optimization problem in (\ref{eq:opt}) for the next $\omega$ time slots, \ie over the window of $(t,t+\omega)$ given the initial battery level $b_{i}^{t-1}$, $\forall i \in \Nset$ and estimated parameters set $\boldsymbol z^{t:t+\omega}$. This will give us $\boldsymbol x^{t}, \ldots, \boldsymbol x^{t+\omega}$, which are interpreted as the set of decision policies for the next $\omega$ slots. Then, only the first value of this trajectory is employed, \ie $\boldsymbol x^{t}$. Similarly, at time slot $t+1$, the problem in (\ref{eq:opt}) is solved, we get the results from  $\boldsymbol x^{t+1}, \ldots, \boldsymbol x^{t+\omega+1}$, but we only use $\boldsymbol x^{t+1}$ as a solution at $t+1$.  

The rationale behind solving (\ref{eq:opt}) over a time horizon in AACT is to take into account upcoming (in a window) device and application constraints and requirements. If this information is not available for a large time horizon, the window size $w$ could be decreased. Thus, $w$ could be dynamically arranged depending on how much future data is available in the cooperative system. 

We note that  AACT is centralized, so one of the cooperating IoT devices behave as a central node, and make all decisions. Next, we present the distributed version of this algorithm; \ie AACT-distributed.


\subsection{AACT-distributed}


At each time slot $t$, one of the devices is randomly selected as the decision maker (\eg $i$). The selected device and all other devices in the system communicate to exchange the decision parameters such as $\bar{w}_{j}^{t},\bar{c}_{j}^{t}$ and $b_{j}^{t-1} \forall j\neq i \in \mathcal{N}$. Furthermore, since decisions are made according to RHC approach, the values of these parameters must be known for a window time of size $w$. However, this may introduce too much overhead on the system, and the decision parameters may not be exchanged in a timely manner. Thus, selected device (device $i$) sets
\begin{align}
\bar{c}_{j}^{t:t+w}=\bar{c}_{j}^{t}, \mbox{   } \bar{w}_{j}^{t:t+w}=\bar{w}_{j}^{t}, \forall j\neq i \in \mathcal{N}, 
\end{align} which means that device $i$ only uses the decision parameters at time $t$ for all neighboring devices. Yet, for its own parameters, it knows all the decision variables in a time window $(t,t+\omega)$ and uses these values. Our AACT-distributed algorithm is summarized in Algorithm~\ref{alg:distributed}.

\begin{algorithm}
\caption{AACT-distributed}\label{alg:distributed}
\begin{algorithmic}[1]
\State for $t\in \mathcal{T}$ do
\State $\quad$ Select one of the mobile devices randomly (e.g., $i$)
\State $\quad$ for $j\in \mathcal{N},j\neq i$ do
\State $\quad$ $\quad$ $\bar{c}_{j}^{t:t+w}=\bar{c}_{j}^{t}$
\State $\quad$ $\quad$ $\bar{w}_{j}^{t:t+w}=\bar{w}_{j}^{t}$
\State $\quad$ Solve optimization problem in (\ref{eq:opt}).
\end{algorithmic}
\end{algorithm}

%% file: simulation.tex
\section{\label{sec:sims} Performance Evaluation}

\subsection{Setup} In this section, we evaluate the performance of our algorithm; AACT as compared to two baselines; No-cooperation and Cooperation (sequential). In No-cooperation, each device runs its own application without any cooperation. In Cooperation (sequential), all devices cooperate, but applications are handled in a sequence. For example, if there are two applications such as video streaming and puzzle solving, this baseline randomly selects one of the applications, runs the selected application first cooperatively among all the devices, and then it runs the second application. 

We consider a topology shown in Fig.~\ref{fig:example} for a different number of devices and heterogeneous and time varying device constraints and application requirements. Optimization time frame $T$ and window size $w$ vary depending on simulation scenario. We consider two applications: $a_{1}$, which is a video streaming application, and $a_{2}$, which is a computationally intensive application such as a game or puzzle. The resource requirements of applications are $a_{1}^{c}=0.5 Ghz, a_{2}^{c}=1.5 Ghz, a_{1}^{w}=1 Mb/sec$ and $a_{2}^{w}=0 Mb/sec$ (\ie no Internet connection is needed for $a_2$). Device specifications vary depending on the simulation scenario. 

We consider two performance metrics for our evaluation: (i) completion time $T_{j}$ of application $j$, defined as the number of time slots needed to process the application, and (ii) battery consumption of each device.

\subsection{Simulation Results}

\begin{figure}[t!]
\centering
\vspace{-5pt}
\subfigure[AACT]{ {\includegraphics[height = 30mm]{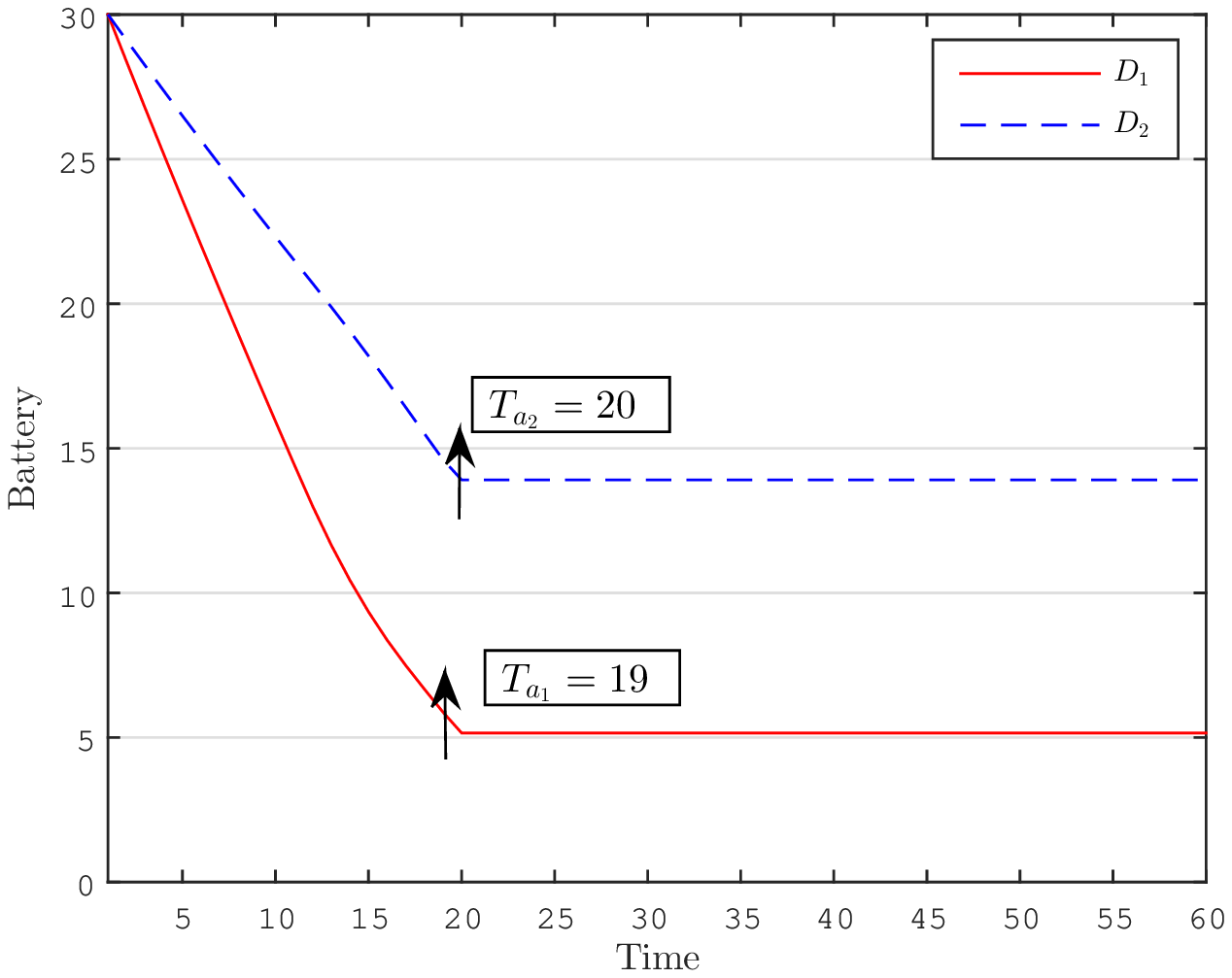}} } 
\subfigure[No-cooperation]{ {\includegraphics[height = 30mm]{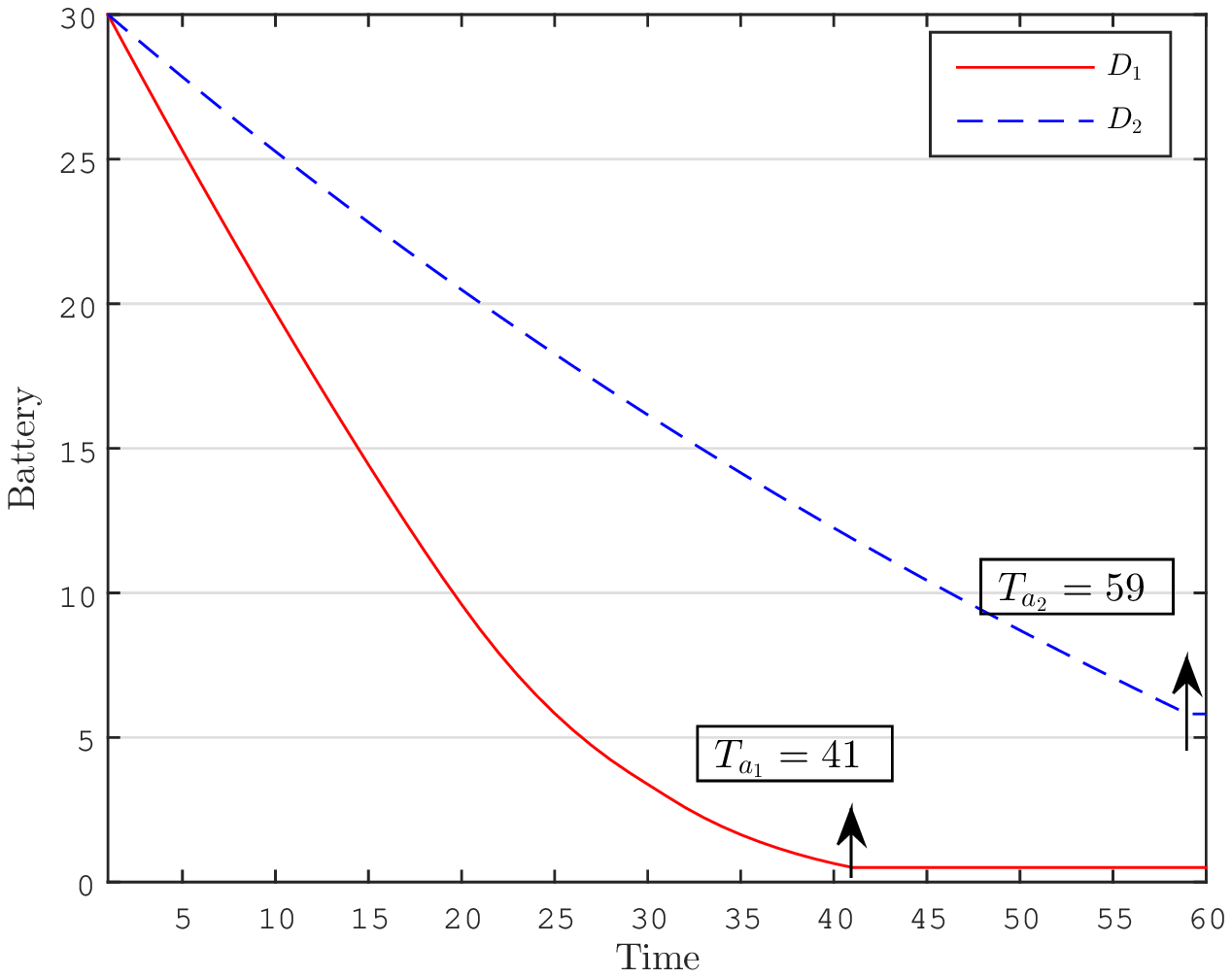}} } 
\vspace{-5pt}
\caption{Battery consumption of each device over time for AACT and No-cooperation.The completion time of each application is marked with a black arrow. 
}
\vspace{-10pt}
\label{fig:Baseline1}
\end{figure}

In the first scenario, we consider that the device specifications are set to $\bar{c}_{1}=1 GHz,\bar{c}_{2}=0.5 Ghz$ and $\bar{w}_{1}=0.25 Mbps,\bar{w}_{2}=0.5 Mbps$, and does not vary over time. The total time frame and the window size used are set to $T=60$ steps and $\omega=10$, respectively. Fig. \ref{fig:Baseline1} demonstrates the battery level of each device over time for AACT and No-cooperation. In the same figure, black arrows show the completion times of each application. As expected, the completion times are smaller in the cooperative set up ($T_{a_{1}}=19$,$T_{a_{2}}=20$) as compared to the No-cooperation set up ($T_{a_{1}}=41$,$T_{a_{2}}=59$) thanks to cooperatively using available resources. 
Furthermore, AACT significantly improves battery usage as compared to No-cooperation.

\begin{figure}[t!]
\centering
\vspace{-5pt}
\subfigure[AACT]{ {\includegraphics[height = 30mm]{coop.eps}} } 
\subfigure[Cooperation (sequential)]{ {\includegraphics[height = 30mm]{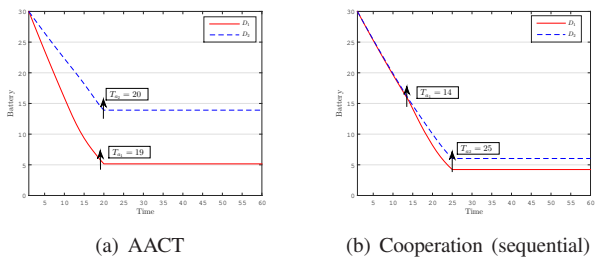}} } 
\vspace{-5pt}
\caption{Battery consumption of each device over time for AACT and Cooperation (sequential).The completion time of each application is marked with a black arrow. 
}
\vspace{-15pt}
\label{fig:Sequential}
\end{figure}

For the same set up, we compare AACT and Cooperation (sequential) in Fig. \ref{fig:Sequential}. As seen, Cooperation (sequential) finishes the first application earlier than AACT as application one is chosen to be processed first by the cooperative set up. The overall completion time of applications in AACT is smaller than Cooperation (sequential) and there is a significant improvement in terms of the energy consumption in AACT. This is reasonable as the sequential process of applications is a feasible solution, but not necessarily optimal. Note that AACT can run applications in any order.

Now, we consider the impact of number of devices on the completion time of AACT and Cooperation (sequential). Note that we do not consider No-cooperation in this scenario as the number of devices does not affect the performance of No-cooperation.  The completion times are presented in Tables \ref{tab:aact} and \ref{tab:seq} for AACT and Cooperation (sequential), respectively. As seen, AACT improves the total completion time of applications as compared to Cooperation (sequential). When the number of devices increases, the total completion time reduces, but AACT still improves it as compared to Cooperation (sequential). 


\begin{table}[h!]
 \caption{Completion time of AACT}
 \centering
 \begin{tabular}{|c|c|c|}
\hline
Number of devices & Application one & Application two \tabularnewline
\hline
$N=1$  & $14$ & $19$\tabularnewline
\hline
$N=2$  & $6$ & $7$\tabularnewline
\hline
$N=3$  & $4$ & $4$\tabularnewline
\hline
$N=4$  & $3$ & $3$\tabularnewline
\hline
 \end{tabular}
\label{tab:aact}
\end{table}

\begin{table}[h!]
\caption{Completion time of Cooperation (sequential)}
\centering
\begin{tabular}{|c|c|c|}
\hline
Number of devices & Application one & Application two \tabularnewline
\hline
$N=1$  & $11$ & $30$\tabularnewline
\hline
$N=2$  & $5$ & $11$\tabularnewline
\hline
$N=3$  & $3$ & $7$\tabularnewline
\hline
$N=4$  & $2$ & $5$\tabularnewline
\hline
 \end{tabular}
\label{tab:seq}
\end{table}

\begin{figure}[t!]
\centering
\vspace{-5pt}
\subfigure[Completion time]{ {\includegraphics[height = 30mm]{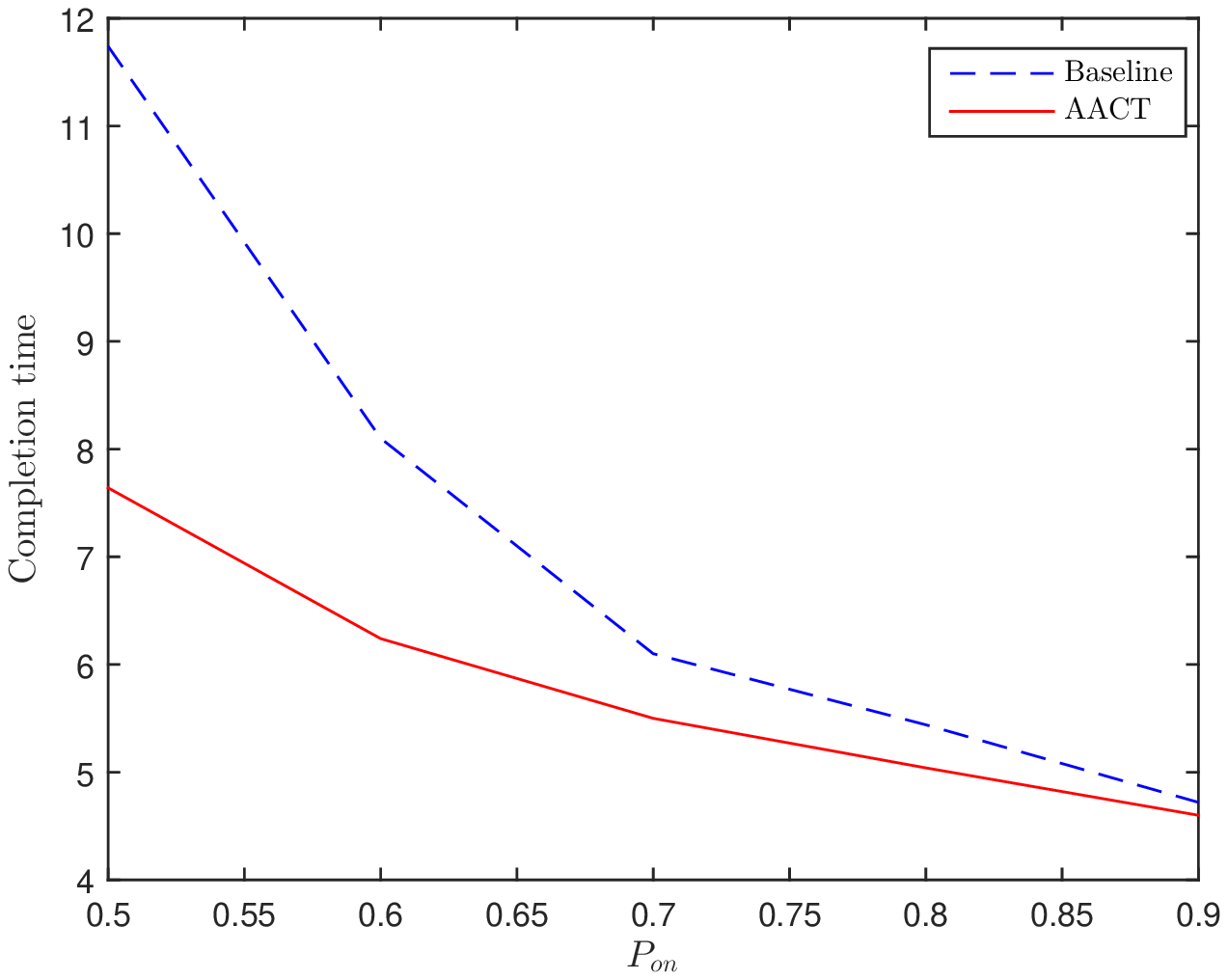}} } 
\subfigure[Battery ]{ {\includegraphics[height = 30mm]{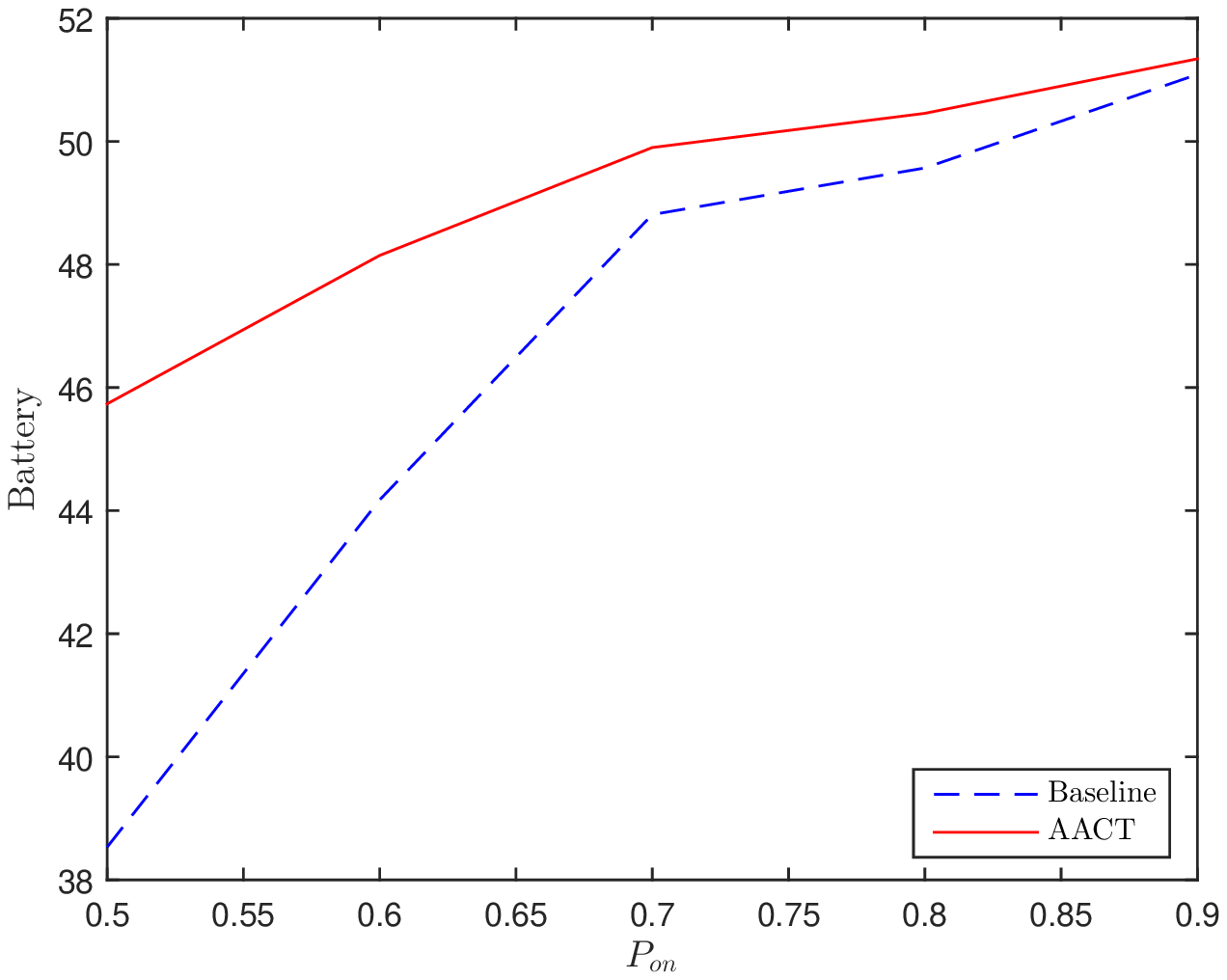}} } 
\vspace{-5pt}
\caption{Completion time of application one (video streaming) along with the remaining battery level at the completion time is plotted vs $P_{on}$ in (a) and (b). Device specifications are $\bar{c}_{1}=1,\bar{c}_{2}=0.5Ghz$, $\bar{w}_{1}=0.5,\bar{w}_{2}=0.25 Mb/sec$ and $\bar{S}_{a_{1}}=2,\bar{S}_{a_{2}}=5$.
}
\vspace{-5pt}
\label{fig:OnOff1independent}
\end{figure}

Now, we consider that device parameters vary over time. In particular, cellular bandwidth follows a Bernoulli ON/OFF distribution. We assume that the downlink cellular channel is ON with a certain probability denoted as $P_{on}$ and it is off with the probability $1-P_{on}$. In this scenario, our goal is to investigate the benefit of RHC in our AACT algorithm. We first consider AACT-oracle, where we know exactly if channel will be ON or OFF for the next $\omega$ time slots. We also consider AACT-average, where we know average values of ON/OFF probabilities. We set $T=300$, windows size to $\omega=20$. The other device specifications are as follows: $\bar{c}_{1}=1,\bar{c}_{2}=0.5Ghz$, $\bar{w}_{1}=0.5,\bar{w}_{2}=0.25 Mb/sec$. Fig~\ref{fig:OnOff1independent} demonstrates a significant improvement (in terms of the completion time of application and energy consumption) of AACT-oracle as compared to AACT-average thanks to having exact knowledge about channel conditions in a time window, and using this information efficiently using an RHC approach.

\begin{figure}[t!]
\centering
\vspace{-5pt}
\subfigure[Completion time]{ {\includegraphics[height = 30mm]{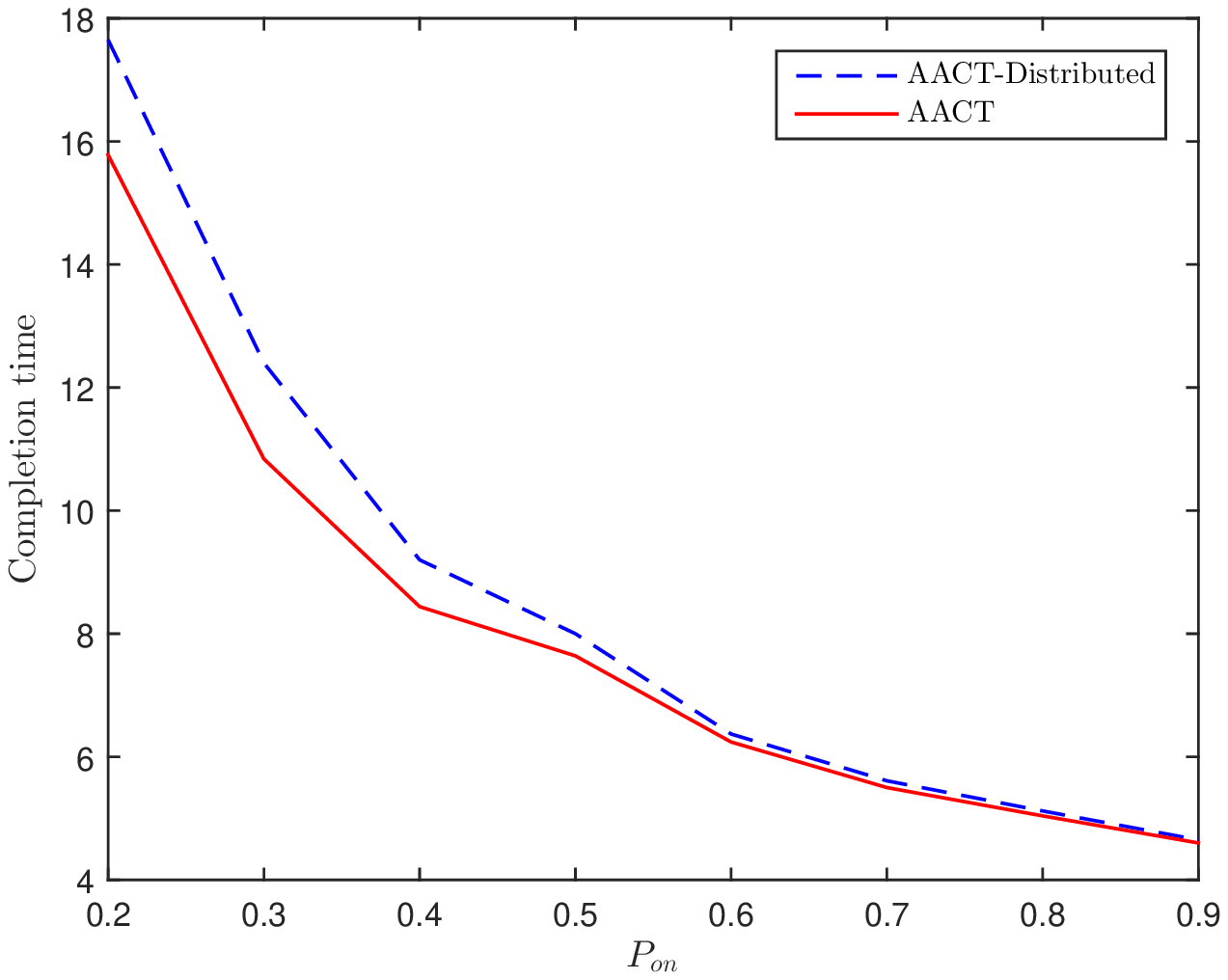}} } 
\subfigure[Battery]{ {\includegraphics[height = 30mm]{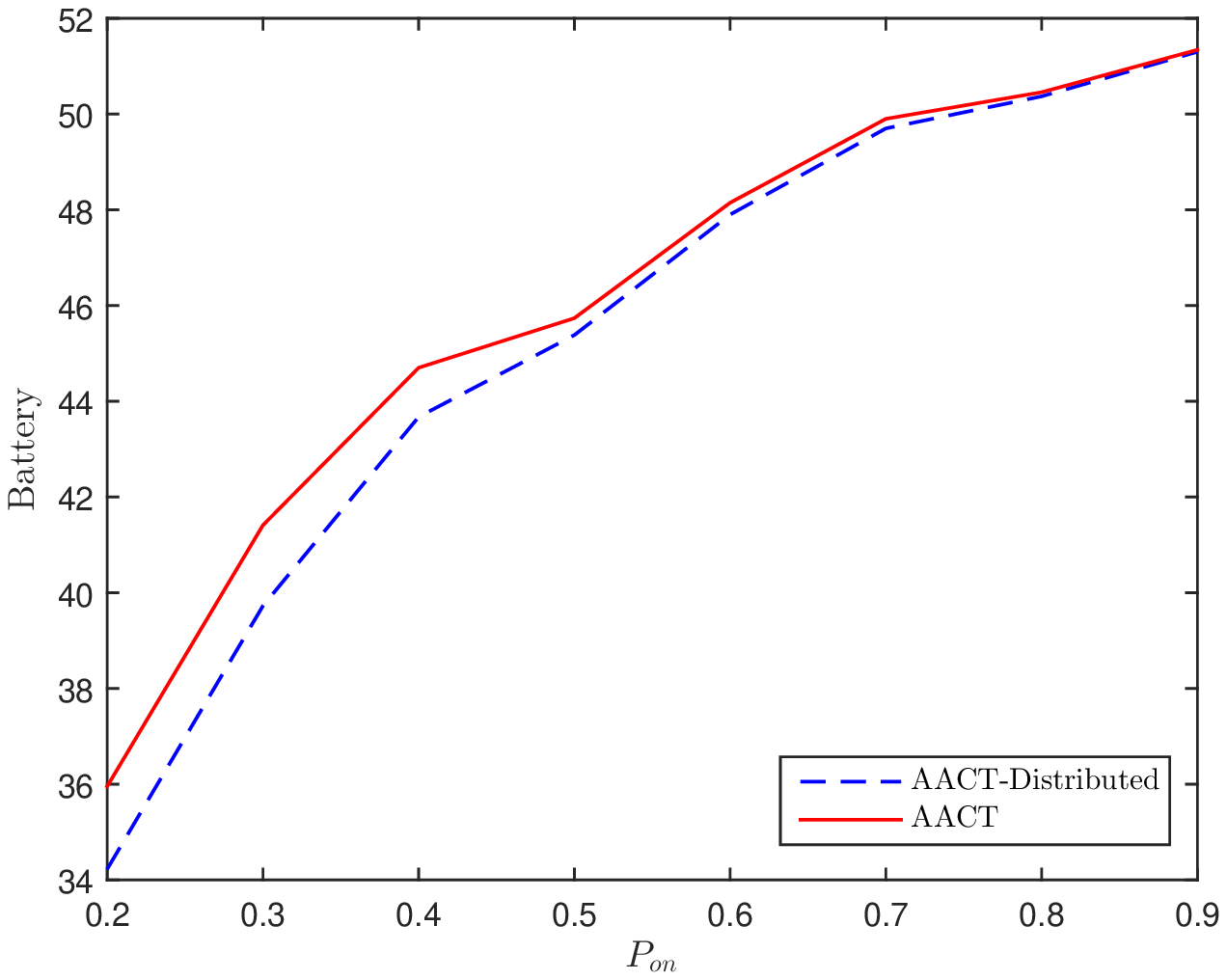}} } 
\vspace{-5pt}
\caption{Completion time of application 1 (video streaming) along with the remaining battery level at the completion time is plotted vs $P_{on}$ in (a) and (b), respectively. Device specifications are $\bar{c}_{1}=1,\bar{c}_{2}=0.5Ghz$, $\bar{w}_{1}=0.5,\bar{w}_{2}=0.25 Mb/sec$ and $\bar{S}_{a_{1}}=2,\bar{S}_{a_{2}}=5$.
}
\vspace{-15pt}
\label{fig:OnOffdistributed}
\end{figure}

We consider the same setup, but now we evaluate the performance of AACT-distributed as compared to AACT. Fig. \ref{fig:OnOffdistributed} demonstrates that the completion time and energy consumption performance of AACT-distributed is very close to its centralized version (AACT), which shows the effectiveness of our distributed algorithm.

%% file: related.tex
\section{Related Work}
\label{sec:related}

Cooperation among IoT devices by exploiting D2D connections is getting increasing interest, where aggregated and complementary resources in a cooperative setup can be utilized by all devices. Such a cooperative system has been widely considered for throughput improvement and cellular offloading, cooperative computation (or computation offloading), and energy efficiency. 


D2D networking has been widely used for the purpose of throughput enhancement in cooperative data streaming in wireless mobile networks \cite{keller2012microcast,gong2016design,li2016incentivizing}. Authors in \cite{keller2012microcast} and \cite{gong2016design} designed a cooperative video streaming system to enhance the streaming experience in a group of mobile device users by utilizing both cellular and Wifi interface of devices. The work in \cite{li2016incentivizing}  develops an incentive framework to promote the cooperation for streaming a live content in a group of co-located wireless devices. As compared to this line of work, we consider multiple applications with heterogeneous resource requirements, and developed a resource allocation mechanism for this setup.

Computation offloading (as also called mobile edge computing) gained a significant interest in academia in recent years and is known to be a promising technique to mitigate the limited computing capability of the mobile devices \cite{mao2016dynamic,chen2017computation}. In this framework, the computation tasks are offloaded to a physically proximal MEC server in order to reduce the computation workloads on the local device and hence enhance the quality of computation experience. The works in \cite{mao2016dynamic} and \cite{chen2017computation} proposed a peer-offloading framework to optimize edge computing performance by taking into account the limited energy resources. Nonetheless, the efficiency of computation offloading in MEC is greatly influenced by the condition of channel as the data must be transmitted between mobile devices and MEC servers \cite{mao2016dynamic}. The work in \cite{kim2015dual} addressed this issue by proposing dual-side control algorithms which jointly optimizes both the user-side and cloud-side under the non-static assumption of wireless channel. As compared to this line of work, we consider other applications in addition to computation applications, where different applications can be simultaneously run by cooperating IoT devices.


Numerous work has been done from the perspective of energy efficiency in D2D wireless networks \cite
{ramadan2008implementation,shen2016device,singh2016energy}. Energy-aware cooperative frameworks were developed for a group of connected D2D devices engaged in video streaming \cite{singh2016energy,shen2016device} and computationally intensive tasks \cite{ramadan2008implementation}. As compared to this line of work we optimize the time utilized by applications by taking into account different device specifications including energy, computing power as well as the cellular rates.

Receding Horizon Control (RHC) has received significant interest especially from the control theory community as well as networking \cite{bartolini2013thermal,quevedo2015stochastic,mao2016autonomous}. MPC is an effective tool to deal with multivariable constrained optimization problems under uncertainty. In this work, we benefit from RHC framework to take into account the downlink channel variations to make optimal resource allocation decisions. As compared to existing work, we apply RHC to solve our application-aware cooperative time allocation problem.

%% file: conclusion.tex
\section{Conclusion}
\label{sec:conclusion}

In this paper, we considered a group of cooperative IoT devices with multiple heterogeneous applications and developed Application-Aware Cooperative Time allocation (AACT) framework, which optimizes the time that each application utilizes the aggregated system resources by taking into account heterogeneous device constraints and application requirements. The proposed framework was grounded on RHC where decisions are made by repeatedly solving a convex optimization problem over a moving control window of estimated system parameters. Furthermore, a distributed version of AACT was proposed enabling the devices to cooperatively making the time allocations decisions. Simulation results show that the proposed framework improves the application completion times as well as the energy consumption of devices. 

%% file: refs.tex
\bibliographystyle{IEEEtran}